# Exploratory Factory Analysis of the Centrality Metrics for Complex Real-World Networks

Natarajan Meghanathan
Professor of Computer Science
Jackson State University, Jackson MS, USA
Email: natarajan.meghanathan@jsums.edu

**Abstract**
Exploratory factor analysis (EFA) is useful to identify the number and mapping of the hidden factors that could dominantly represent the features in the dataset. Principal component analysis (PCA) is the first step as part of the two-step procedure to conduct EFA, with the number of dominant principal components being the number of hidden factors and the entries for the features in the corresponding Eigenvectors serve as the initial values of the factor loadings. In this paper, we conduct EFA on a suite of 80 complex network datasets to identify the number and mapping of the hidden factors (expected to be less than four) that could dominantly represent the values incurred by the vertices with respect to the four major centrality metrics (degree: DEG, eigenvector: EVC, betweenness: BWC and closeness: CLC).

## 1 Introduction

Centrality metrics quantify the topological importance of nodes or edges in a complex network [1]. The four major centrality metrics typically studied for complex network analysis are: degree centrality (DEG), eigenvector centrality (EVC), betweenness centrality (BWC) and closeness centrality (CLC). The degree centrality [1] of a node is a direct measure of the number of neighbors of the node. The eigenvector centrality [2] of a node is a measure of the degree of the node as well as the degrees of its neighbors. The betweenness centrality [3] of a node is a measure of the fraction of the shortest paths between any two nodes in the network that go through the node. The closeness centrality [4] of a node is a measure of the closeness of the node to the rest of the nodes in the network. Among these centrality metrics, the degree centrality metric is computationally-light and can be computed without requiring global knowledge. Whereas, the other three major centrality metrics (EVC, BWC and CLC) are computationally-heavy and require global knowledge.

We propose to conduct Exploratory Factor Analysis (EFA [11-12]) on a complex network dataset featuring the values incurred by the vertices with respect to the four major centrality metrics (DEG, EVC, BWC and CLC) and identify the number of hidden factors (expected to be less than 4, the number of centrality metrics considered for EFA) that could dominantly represent the values incurred for the four centrality metrics as well as extract a mapping of the dominating factors to the centrality metrics represented. Through such an EFA, we seek to assess whether the two neighborhood-based centrality metrics (DEG and EVC) are dominantly represented by the same factor and likewise, whether the two shortest paths-based centrality metrics (BWC and CLC) are dominantly represented by the same factor. In this context, we also evaluate the canonical correlation coefficient (CCA) between the (DEG, EVC) vs. (BWC, CLC) sets of centrality metrics and quantitatively evaluate the impact of the strength of the canonical correlation between these two sets of centrality metrics on the results of EFA.

The rest of the paper is organized as follows: Section 2 presents the step-by-step procedure to conduct exploratory factor analysis on a dataset of the centrality metrics for a toy example graph as well as the procedure to extract the mapping of the dominating factors to the centrality metrics that they could represent. Section 3 presents the (DEG, EVC) vs. (BWC, CLC) canonical correlation coefficient values obtained for a suite of 80 real-world networks that are also used for the EFA study. Section 4 presents the results of EFA (the number of hidden factors and the mapping of the dominating factors to the centrality metrics) conducted on the same suite of 80 real-world networks. Section 5 reviews related work in the literature and highlights our contributions. Section 6 concludes the paper and presents plans for future work. Throughout the paper, the terms 'network' and 'graph', 'node' and 'vertex', 'edge' and 'link' are used interchangeably. They mean the same.

## 2 Exploratory Factor Analysis of Centrality Dataset

Exploratory Factor Analysis (EFA) [11-12] is a widely used approach in machine learning to quantify the hidden factors that are behind the values incurred for the different features (columns) in a dataset (matrix) of records. For the problem in hand, the features are the four centrality metrics (DEG, EVC, BWC and CLC) and the records are the centrality metric values incurred for the vertices. We now explain the procedure to conduct EFA using a toy example graph. We use the normalized centrality metric values for the vertices as the dataset and first determine its covariance matrix (see Figure 1). The covariance matrix comprises of the Pearson's correlation coefficient [5] between the centrality metrics values for any two nodes.

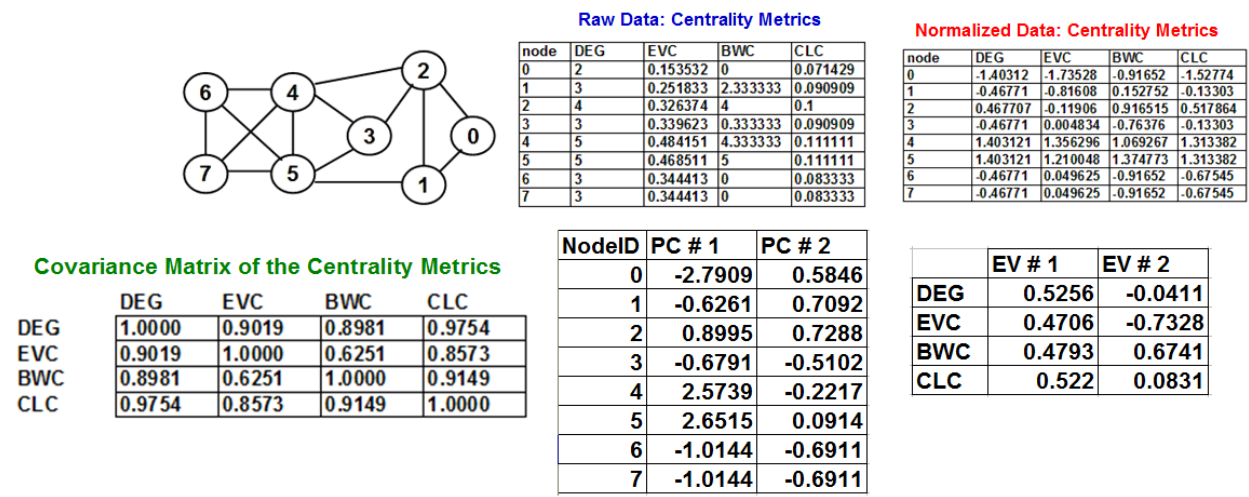

**Figure 1:** Running Example Graph; Centrality Metrics Dataset; Covariance Matrix and its Principal Components Retained and their Corresponding Eigenvectors

We then determine the Eigenvalues and Eigenvectors [5] of the Covariance matrix. Since the Covariance matrix is a 4 x 4 matrix (corresponding to the number of centrality metrics considered), we will get four Eigenvalues and the corresponding Eigenvectors. We multiply the normalized dataset of the centrality metrics with each of the four Eigenvectors to obtain the corresponding principal components. We compute the variances of the entries in each of the principal components and retain the first *m* Eigenvectors whose sum of the variances is at least 99% of the number of rows/columns of the Covariance matrix. The number of Eigenvectors retained correspond to the number of hidden *factors* that could dominantly represent the features of the dataset. Note that the entries in the principal components correspond to the node ids and the entries in the Eigenvectors correspond to the features (centrality metrics in our case). We refer to the communality score [14] for each entry (centrality metric) in these Eigenvectors as the sum of the squares of the factor loadings for the centrality metric.

In the case of the toy example graph, we observe the sum of the variances (3.5950 and 0.3784) of the first two principal components is 99.33% of 4 (the number of rows/columns of the Covariance matrix) and hence we retain the Eigenvectors corresponding to these two principal components (all of which are shown in Figure 1). The number of Eigenvectors retained (two, for the toy example graph) correspond to the number of hidden factors in the dataset and we consider the entries in these Eigenvectors as the initial values of the factor loadings (a measure of the extent of the representation of the centrality metrics by these factors). The communality score for the DEG metric based on the initial factor loadings is $(0.5256)^2$

+ $(-0.0411)^2$ = 0.2779. Likewise, the communality scores for the EVC, BWC and CLC metric based on the initial factor loadings are $(0.4706)^2$ + $(-0.7328)^2$ = 0.7585, $(0.4793)^2$ + $(-0.6741)^2$ = 0.6841, $(0.5220)^2$ + $(0.0831)^2$ = 0.2794 respectively.

The communality score (whose maximum value can be 1.0) for a feature (centrality metric in our case) is a quantitative measure of the extent of representation of the feature in the identified factors. Varimax rotation [15] is a coordinate rotation procedure to scale the loadings for the features (centrality metrics) with respect to the factors such that the communality scores for each of the features (centrality metrics) is maximized (in our case, at least 0.98). In order to perform Varimax rotation, we first plot the initial factor loadings for the centrality metrics in a coordinate system (for the toy example graph, we would need a 2-dimensional coordinate system as EFA identified two factors to represent the four centrality metrics) referred to as the principal Eigenvector-axes; see the 2-dimensional coordinate system to the left of Figure 2. Varimax rotation involves repeated orthogonal synchronous rotations of the Eigenvector coordinate-axes until the sum of the squares of the loadings for each feature is maximized. We conducted Varimax rotation using the relevant libraries available in Python (Pandas [16]). The axes in the resulting rotated coordinate system (see to the right of Figure 2) correspond to the two factors and the coordinates for the features (centrality metrics) in this coordinate system represent the final factor loadings for the features.

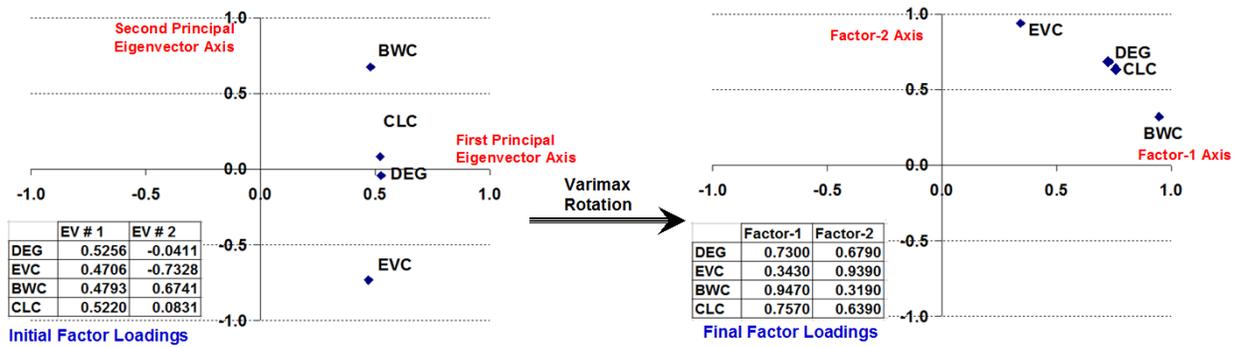

**Figure 2:** Varimax Rotation of the Eigenvector-Axes to the Factor-Axes

The final factor loadings for the four centrality metrics resulting from Varimax rotation (mentioned in the right of Figure 2) are: DEG (0.7300, 0.6790), EVC (0.3430, 0.9390), BWC (0.9740, 0.3190) and CLC (0.7570, 0.6390). Based on these final factor loadings, we observe the communality scores for each of DEG, EVC and BWC to be greater than 0.99 and the communality score for CLC to be greater than 0.98. To identify the factor that could be considered to dominantly represent a centrality metric, we pick the factor whose entry in the final factor loading tuple is the maximum. In case of a tie, we consider the metric to be represented by all the tying factors. Note that by conducting Varimax rotation until the communality score for a centrality metric is at least 0.98, we are requiring that more than 50% (and more than 70%) of the centrality metric is represented by one of the three factors (and two factors) in a three-factor (and two-factor) coordinate system. On this basis, for the toy example graph, the DEG (0.7300), BWC (0.9470) and CLC (0.7570) metrics are dominated by Factor-1 and the EVC (0.9390) metric is dominated by Factor-2. Thus, for the toy example graph, we conclude the two shortest paths-based centrality metrics are dominantly represented by the same factor, whereas the two neighborhood-based centrality metrics are dominantly represented by different factors.

## 3   Canonical Correlation Analysis for Real-World Network Graphs
Canonical correlation analysis (CCA) is conducted to quantitatively assess the strength of correlation between two sets of metrics, in this paper the (DEG, EVC)-neighborhood based centrality metrics vs. the (BWC, CLC)-shortest paths based centrality metrics. Like Pearson's correlation coefficient (used to assess the strength of correlation between two individual metrics), the canonical correlation coefficient (CCC) values also range from -1 to 1: positive CCA values indicate the values incurred for one set of metrics increase (decrease) with increase (decrease) in the values incurred for the other set of metrics; negative CCC values indicate the values incurred for one set of metrics increase (decrease) with decrease (increase) in the values incurred for the other of metrics. We seek to the impact of the strength of the canonical correlation between the two categories of centrality metrics on the representation of each of the two categories of centrality metrics by the same factor or different factors.

In this section, we present results of the canonical correlation analysis [21] conducted on a suite of 80 real-world network graphs whose number of nodes ranges from 22 to 1,538 (with a median of 143 nodes) and the number of edges ranges from 38 to 16,715 (with a median of 613 edges). We computed the values of the four major centrality metrics (DEG, EVC, BWC and CLC) of the nodes in each of these networks using the appropriate algorithms mentioned earlier. We compute the canonical correlation coefficient (CCC) values for the neighborhood-based (DEG, EVC) metrics vs. the shortest path-based (BWC, CLC) metrics.

Figure 3 presents the distribution (sorted in the decreasing order) of the CCC values for the 80 real-world networks. We observe 30 and 50 of the 80 real-world networks to respectively incur negative and positive CCC values. While the negative CCC values fall in the range of [-1, ..., -0.79]: indicating a strong negative correlation, the positive CCC values fall in the range of [0.3, ..., 1]. Among the 50 real-world networks incurring positive CCC values, 37 of them incur values above 0.79, indicating a strong positive correlation. Only one of the 80 real-world networks incurred a CCC in the range of (-0.79, ..., 0.53). Overall, 67 of the 80 real-world networks (i.e., more than 4/5th of the 80 real-world networks) are observed to exhibit either a strong positive or strong negative canonical correlation between the neighborhood-based centrality metrics and the shortest path-based centrality metrics. This also implies that the ranking of vertices with respect to one category of centrality metrics is more likely to be the same (for strongly positive canonical correlated networks) or more likely the opposite (for strongly negative canonical correlated networks) as the ranking of vertices with respect to the other category of centrality metrics.

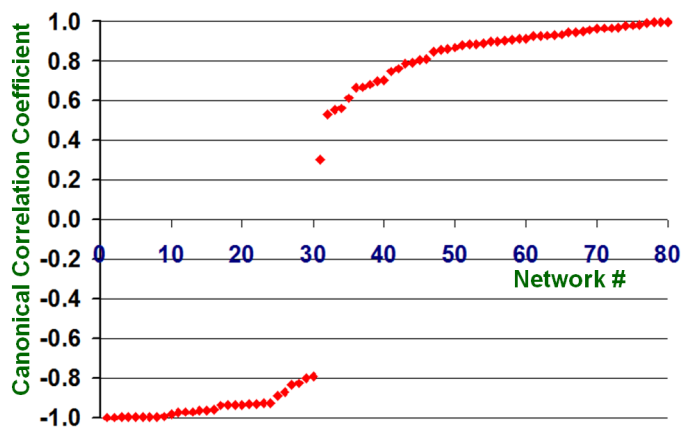

**Figure 3:** Canonical Correlation Coefficient Values of the Real-World Networks

## 4 Exploratory Factor Analysis for the Real-World Network Graphs

We now present results of Exploratory Factor Analysis (EFA) conducted on the same suite of 80 real-world network graphs for which we conducted CCA and presented the CCC results in Section 3. We seek to identify the number of hidden factors behind the values incurred by the vertices in these 80 real-world networks with respect to the four centrality metrics DEG, EVC, BWC and CLC as well as extract a mapping of the factors to these centrality metrics. We want to explore whether the two neighborhood-based centrality metrics DEG and EVC get mapped to the same dominating factor as well as whether the two shortest paths-based centrality metrics BWC and CLC get mapped to the same dominating factor. As the number of features in the dataset is 4, we restrict our EFA to exploring whether one, two or three hidden factors might be behind the values incurred for the centrality metrics.

Overall, we observe two factors to be sufficient to statistically represent (with a communality score of 0.98 or above for each metric) the four centrality metrics for networks that exhibited either a strong positive or negative canonical correlation [21] between the neighborhood-based and shortest paths-based centrality metrics. On the other hand, we observe three factors are needed to statistically represent the four centrality metrics for networks that exhibited a weak-moderate canonical correlation between these two categories of centrality metrics.

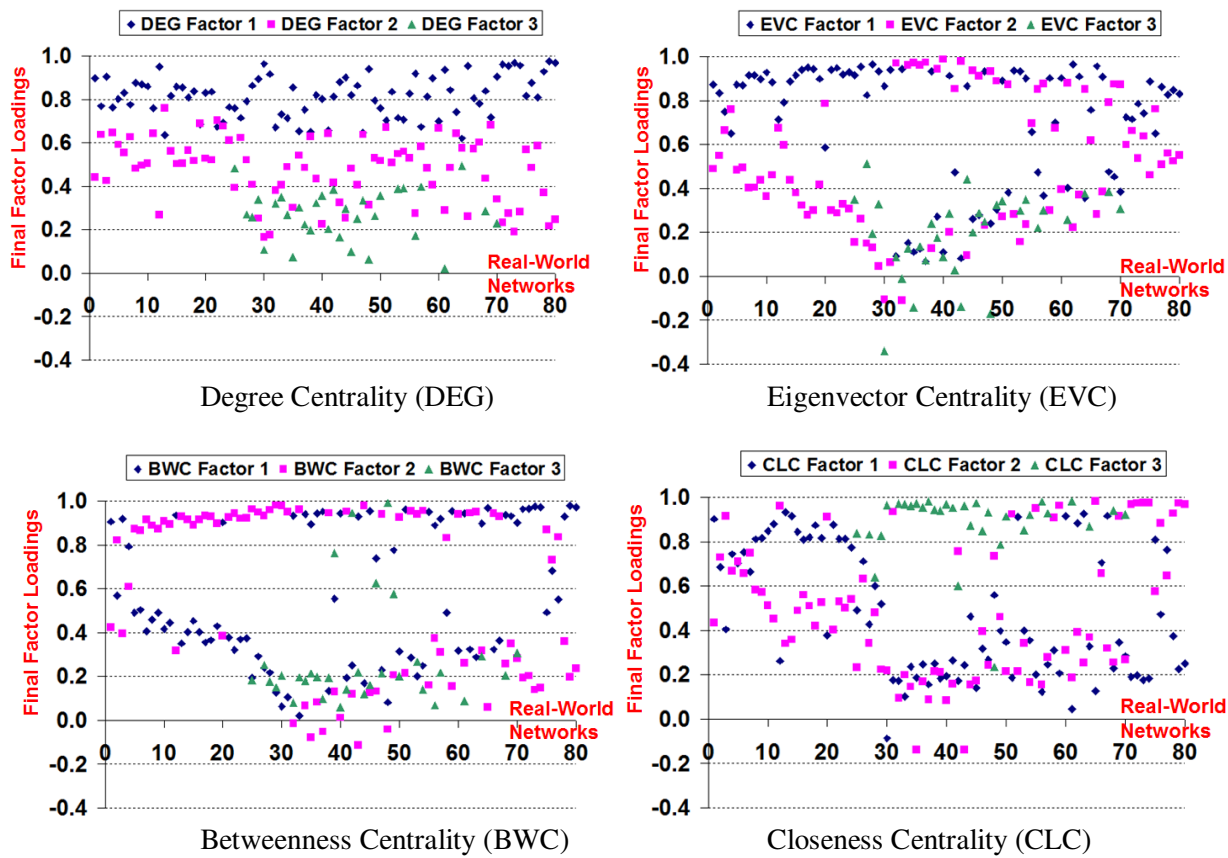

**Figure 4:** Final Factor Loadings for the Centrality Metrics with respect to the 80 Real-World Networks

Figure 4 presents the results (final factor loadings) for the 80 real-world networks with respect to each of the four centrality metrics. The network #s used to represent the real-world networks in both Figures 3 and 4 are the same. We observe the DEG centrality metric to incur its largest loadings in Factor-1 for all the 80 real-world networks (the blue-color symbol corresponding to Factor-1 is above the other two symbols throughout the 80-networks sequence). The EVC metric also incurs its largest loadings in Factor-1 predominantly for networks that exhibit either strong positive or strong negative canonical correlation. Overall we observe DEG and EVC to be dominated by the same factor in 52 of the 80 real-world networks. For about half of the 31 real-world networks that exhibited a weak-moderate canonical correlation between the neighborhood-based and shortest paths-based centrality metrics, we observe the two neighborhood-based DEG and EVC metrics to be dominated by different factors (Factor-1 for DEG and Factor-2 for EVC).

A significant observation from the EFA results presented in Figure 4 is that the BWC and CLC metrics are dominated by different factors for more than 90% (73) of the 80 real-world networks. This could be justified by the totally different underlying characteristics the two metrics tend to capture. The BWC metric tends to identify nodes through which majority of the shortest paths communication could occur; on the other hand, the CLC metric tends to identify nodes that are geographically closer to the rest of the nodes in the network. Our EFA results show that these two characteristics are indeed different and need not be represented by the same factor.

In relation to the CCA results of Section 3, for a majority of the networks that exhibited strong negative (positive) canonical correlation between the neighborhood-based and shortest paths-based centrality metrics, we observe the CLC (BWC) metric to be dominated by the same factor (Factor-1) that also dominated the DEG and EVC metrics while the BWC (CLC) metric appears to be dominated by a different factor (Factor-2). Note that for the weak-moderate canonically correlated networks, the CLC metric is dominated by a totally different factor (Factor-3, which mostly dominates only the CLC metric) whereas, we observe the BWC metric to be either dominated by Factor-1 (along with DEG) or Factor-2 (along with EVC). The above discussions relating the CCA and EFA results show that between the two shortest paths-based centrality metrics, the BWC metric is relatively a more dominant metric whose variations vis-a-vis the two neighborhood-based metrics contribute to the overall canonical correlation between the two categories of centrality metrics.

With respect to the extent of domination by a particular factor, we observe the BWC metric to be the metric that is most strongly dominated by a particular factor (the BWC-loadings for the dominating factor is much greater than the BWC-loadings for the non-dominating factors), followed by EVC. Though the loading values for the dominating factor vs. the non-dominating factors are clearly separable for the DEG metric, we observe the DEG metric (and to a certain extent the CLC metric) to be relatively less strongly dominated by a particular factor (especially, for 23 of the 24 strong negative canonical correlated networks, the DEG-loading values for the two factors are above 0.4).

## 5  Related Work

Several works have been reported in the literature featuring correlation analysis involving centrality metrics for real-world network datasets. However, these correlation analyses studies involve only two individual centrality metrics. To the best of our knowledge, other than our earlier work reported in [13], there has been no prior work done on canonical correlation analysis (CCA) involving centrality metrics. Also, ours is the first such work to conduct exploratory factor analysis (EFA) involving the four major centrality metrics (DEG, EVC, BWC and CLC) and study the CCA-EFA results in conjunction. For the rest of this section, we review results from some of the relevant correlation and factor analysis studies recently reported in the literature.

In [6], a strong correlation was observed between the two neighborhood-based DEG and EVC metrics, while only a moderate correlation was observed between the two shortest path-based BWC and CLC metrics (also confirmed in [17]) for real-world network graphs of diverse domains. The Pearson's correlation analysis was used in [6]. In [7], the authors used Spearman's rank-based correlation analysis to quantify the extent to which the computationally-light degree centrality metric can be used to rank the vertices in lieu of the computationally-heavy EVC, BWC and CLC metrics: it was observed that high-degree vertices are more likely to be high-BWC vertices as well and high-degree vertices are more likely closer to the rest of the vertices as well; on the other hand, high-degree vertices were observed to be not necessarily vertices with high EVC. The centrality-based clustering results reported in [8] state that there could be two distinct clusters of nodes: a central core cluster in which the constituent nodes have the largest centrality metric values with respect to several measures and a peripheral cluster in which the constituent nodes have lower centrality metric values. In [9], the authors explored the correlation (with respect to Pearson's, Spearman's and Kendall's correlation analysis) between the recently proposed $O(\log^2 n)$ Game of Thieves (GoT) node centrality metric [10] to the four major centrality metrics for synthetic networks generated using the scale-free, small-world and random network models: a strong negative correlation was observed for GoT vs. DEG and BWC, while a weak negative correlation was observed for GoT vs. CLC.

The following factor analysis studies have been reported in the literature: Among the five centrality metrics (Valued centrality [18], BWC, CLC, Flow Centrality [18] and Jordan centrality [19]) considered in [18], the authors use the number of principal components retained (similar to our approach) as the basis to decide on the number factors and conclude the entries for these centrality metrics in the corresponding Eigenvectors as the factor loadings. On the other hand, we treat the entries in the Eigenvectors as the initial factor loadings for the centrality metrics and conduct Varimax rotation to maximize the communality score as well as use the resulting final factor loadings to identify the dominating factor for each centrality metric and explore these results in the context of strong positive, strong negative and weak-moderate canonically correlated networks. Also, similar to our observation, the authors in [18] also observed the BWC and CLC metrics to be represented by two different factors. Likewise, the work of [20] on graph-theoretical models of agricultural landscapes also observes the BWC and CLC metrics to be represented by two different factors, with a third factor to represent the degree and sub graph centrality metrics.

## 6   Conclusions and Future Work

The results of exploratory factor analysis (EFA) on a suite of 80 real-world networks indicate two hidden factors are sufficient (with a communality score above 0.98) to represent the four centrality metrics (DEG, EVC, BWC and CLC) for both the strong positive and negative canonical correlated networks; whereas, three hidden factors are needed to statistically represent the four centrality metrics for weak-moderate canonical correlated networks. We observe the BWC and CLC metrics (though both are shortest paths-based centrality metrics) to be dominantly represented by two different factors for more than 90% of the networks; whereas, the two neighborhood-based centrality metrics (DEG and EVC) to be dominantly represented by the same factor for both the strong positive and negative canonical correlated networks and by two different factors for the weak-moderate canonical correlated networks. Moreover, the BWC (CLC) metric is dominated by the same factor as that of the DEG and EVC metrics for the strong positive (negative) canonical correlated networks.

As part of future work, we plan to build a linear canonical correlation-based prediction model that would use three of the four centrality metrics (such as the DEG, EVC and CLC) as the independent variables and predict the fourth metric (such as the BWC that could be the most computationally-intensive metric) and compare the accuracy of such a prediction with those of the existing Pearson's correlation-based single variable and multi-variable linear prediction models.

**Acknowledgment**
The work leading to this paper was partly funded through the U.S. National Science Foundation (NSF) grant OAC-1835439. The views and conclusions contained in this paper are those of the authors and do not represent the official policies, either expressed or implied, of the funding agency.